\documentclass[%
  english,
  reprint,
  showpacs,
  amsmath,amssymb,
  aps,
  prl,
]{revtex4-1}

\usepackage{graphicx}
\usepackage[tight]{subfigure}
\usepackage{dcolumn}
\usepackage{bm}
\usepackage[colorlinks,linkcolor={blue},citecolor={blue},urlcolor={blue}]{hyperref}
\usepackage{multirow}
\usepackage{array}
\usepackage{booktabs}
\usepackage{ctable}
\usepackage{upgreek}
\usepackage{epsfig}
\usepackage{mathrsfs}
\usepackage{amssymb}
\usepackage{amsbsy}
\usepackage{color}
\usepackage{cancel}
\usepackage{pifont}
\usepackage{marginnote}
\usepackage{dsfont}
\usepackage{babel}

\newcommand{\up}{\uparrow}
\newcommand{\dn}{\downarrow}
\newcommand{\nnum}{\nonumber}
\newcommand{\dotp}[2]{\mathbf{#1}.\mathbf{#2}}
\renewcommand{\vec}[1]{{\boldsymbol{#1}}}
\newcommand{\sg}{\sigma}
\newcommand{\kv}{{\boldsymbol{k}}}

\newcommand{\bcen}{\begin{center}}
\newcommand{\ecen}{\end{center}}
\newcommand{\btab}{\begin{tabular}}
\newcommand{\etab}{\end{tabular}}
\newcommand{\bdes}{\begin{description}}
\newcommand{\edes}{\end{description}}

\newcommand{\beq}{\begin{equation}}
\newcommand{\eeq}{\end{equation}}
\newcommand{\bea}{\begin{eqnarray}}
\newcommand{\eea}{\end{eqnarray}}

\newcommand{\half}{\frac{1}{2}}
\newcommand{\bary}{\begin{array}}
\newcommand{\eary}{\end{array}}
\newcommand{\benum}{\begin{enumerate}}
\newcommand{\eenum}{\end{enumerate}}
\newcommand{\bitem}{\begin{itemize}}
\newcommand{\eitem}{\end{itemize}}

\newcommand{\bk} { \bm{k} }

\newcommand{\bq} { \bm{q} }
\newcommand{\br} { \boldsymbol{r}}

\newcommand{\dou}{\partial}

\newcommand{\D}[1]{\mbox{d}{#1}}

\newcommand{\mean}[1]{\langle #1 \rangle}
\newcommand{\bra}[1]{{\langle #1 |}}
\newcommand{\ket}[1]{| #1 \rangle}

\newcommand{\Itwo}{{\mathds{1}}}

\newcommand{\cH}{{\cal H}}
\newcommand{\cS}{{\cal S}}

\newcommand{\eqn}[1] {eqn.~(\ref{#1})}
\newcommand{\Eqn}[1] {Eqn.~(\ref{#1})}

\newcommand{\Fig}[1]{Fig.~\ref{#1}}

\newcommand{\myhalffig}{0.475\columnwidth}

\newcommand{\myfighalfwidth}{0.4\columnwidth}
\newcommand{\mylabel}[1]{\label{#1}}

\newcommand{\myonlinecite}[1]{[\onlinecite{#1}]}

\newcommand{\mycite}[1]{\cite{#1}}

\newcommand{\opt}[1]{}

\newcommand{\titlename}{Fermionic Superfluid from a Bilayer Band Insulator in an Optical Lattice}

\begin{document}




\title{\titlename}

\author{Yogeshwar Prasad}\email{ypsaraswat@physics.iisc.ernet.in}
\author{Amal Medhi}\email{amedhi@physics.iisc.ernet.in}
\author{Vijay B.~Shenoy}\email{shenoy@physics.iisc.ernet.in}

\affiliation{Center for Condensed Matter Theory, Indian Institute of Science, 
  Bangalore 560012, India}




\date{\today}

\begin{abstract}  
  We propose a model to realize a fermionic superfluid state in an optical lattice circumventing the cooling problem.
  Our proposal exploits the idea of tuning the interaction   in a characteristically
  low entropy state, a band-insulator in an   optical bilayer system, to obtain a
  superfluid. By performing a   detailed analysis of the model including fluctuations
  and augmented by a   variational quantum Monte Carlo calculations of the ground state,
  we show that the superfluid state obtained has high transition temperature of the order
  of the  hopping energy. Our system is designed to suppress other competing orders
  such as a charge density wave. We suggest a laboratory realization of this model via an
  orthogonally shaken optical lattice bilayer.
\end{abstract}


\maketitle

Quantum emulation of interesting condensed matter Hamiltonians using ultra cold atom systems
holds much promise.\mycite{Ketterle2008,Bloch2008,Giorgini2008,Esslinger2010,Bloch2012}
The pace of experimental progress has been impeded by key problems which include simulation
of electromagnetic (gauge) fields, removal of entropy, etc. While the former has seen a
spectacular recent progress\mycite{Lin2009A, Lin2009B,Lin2011,Wang2012,Cheuk2012}, the long
standing ``cooling problem'' of trapped lattice fermions has been more difficult.\mycite{McKay2011}

The cooling problem has been addressed in various ways. One approach has been to find schemes to
``squeeze out'' the entropy.\mycite{Ho2009a,Ho2009b} Others include exploiting metastable
states\mycite{Rosch2008}, using properties of the states (such as the N\'eel state) to develop
cooling protocols\mycite{Paiva2011} (see ref.~\myonlinecite{McKay2011} for a review).
A recent notable proposal is to use an additional beam that helps to enlarge the region
where a desired state is stabilized.\mycite{Mathy2012} Despite this, to the best of our knowledge,
an interesting many body state such as an anti-ferromagnet is yet to be
realized in an optical lattice, while some signatures of fermionic superfluidity have been reported.\cite{Chin2006}

A different strategy would be to create a characteristically low
entropy state in a large region of the trap, and to tune a parameter
(such as the interaction) that drives this low entropy region into an
interesting many-body state. The desiderata of such a many-body state
include (i) high characteristic temperature scale (ii) stability over
other ``uninteresting'' competing states. Here we suggest the use of a
{\em band insulator}, a characteristically low entropy state in which
we tune an attractive interaction to produce a fermionic
superfluid. Band-insulator superfluid transitions have been
investigated earlier in other
contexts.\mycite{Kohmoto1990,Nozieres1999} In fermionic cold atom
systems, motivated by experimental work cited above\cite{Chin2006},
superfluid-band insulator transition engendered by increasing  the
lattice depth with concomitant multiband effects have been
discussed.\cite{Zhai2007,Moon2007,Burkov2009,Nikolic2010,Nikolic2011}
In contrast to these works, our proposal aims to obtain a superfluid in a deep lattice.

We propose and study a {\em bilayer band insulator} that undergoes a transition to a superfluid upon
tuning an attractive interaction, attaining the above desiderata. The model is designed so that
competing phases, such as charge density wave (CDW), are avoided. This is demonstrated by a detailed
analysis including Gaussian fluctuations, and variational Monte Carlo simulations. We show that a
``high-temperature'' superconducting phase is possible in this system by estimating the
Berezinski-Kosterlitz-Thouless transition temperature $T_{BKT}$.  In a regime of parameters,
the system shows interesting physics such as ``pseudogap phenomenon'', even at high
temperatures. We suggest a possible route to realize this in an optical lattice.

\begin{figure}
  \centering
  \centerline{ \includegraphics[width=\myfighalfwidth]{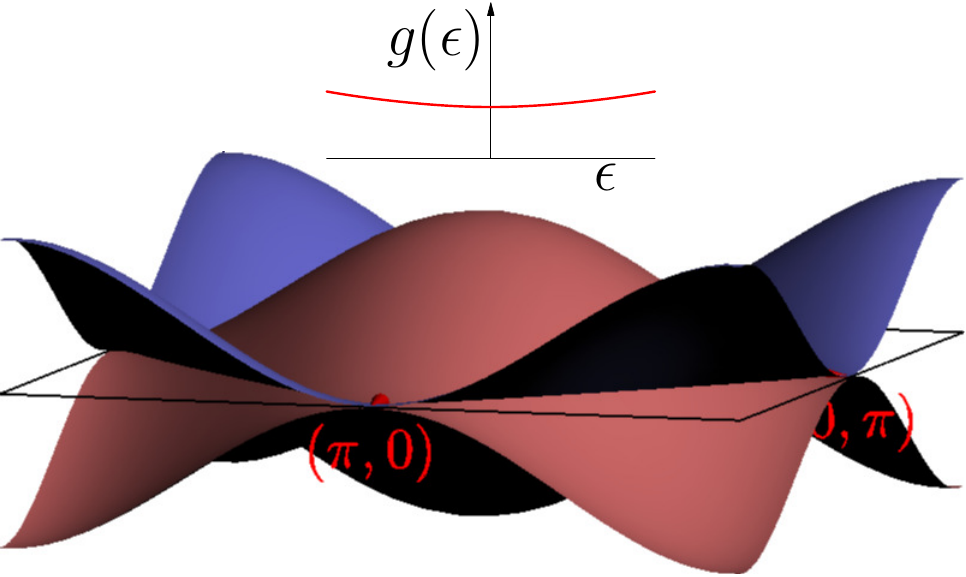} $ \;\;\; \xrightarrow{h(\bk)} \;\;\;$
  \includegraphics[width=\myfighalfwidth]{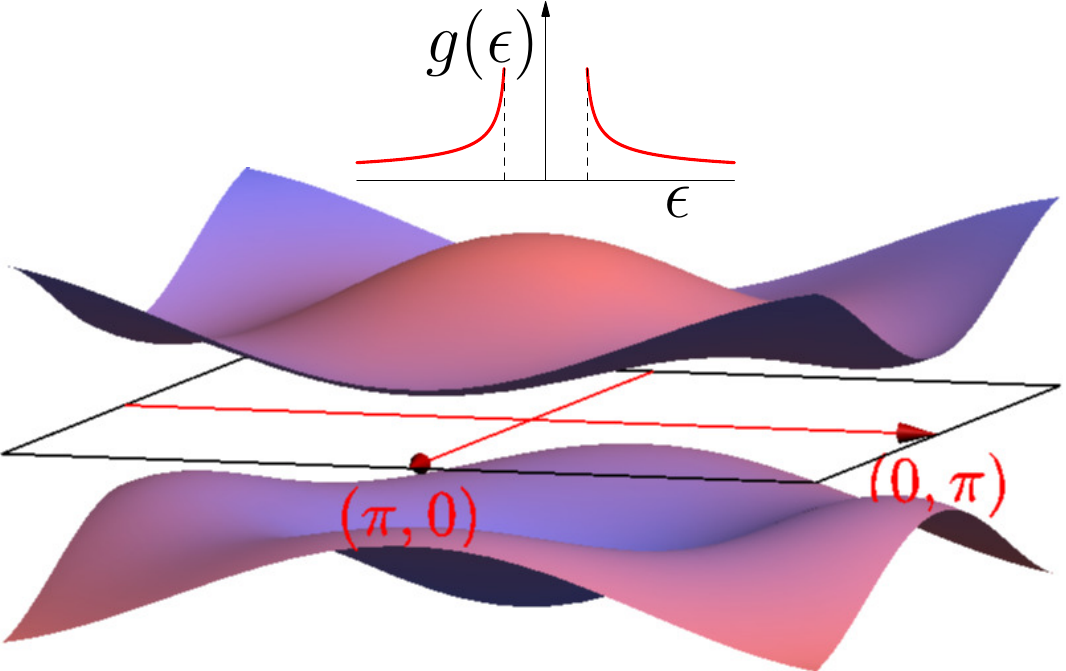} }
  \caption{\bf{Dispersion of the bilayer band insulator} \normalfont -  Bilayer dispersion before (Left) and after (Right)
  interlayer hybridization. Inset show schematic  density of states.}
  \label{fig:scheme}
\end{figure}

\begin{figure}
  \centering 
  \centerline{
     \includegraphics[width=\myhalffig]{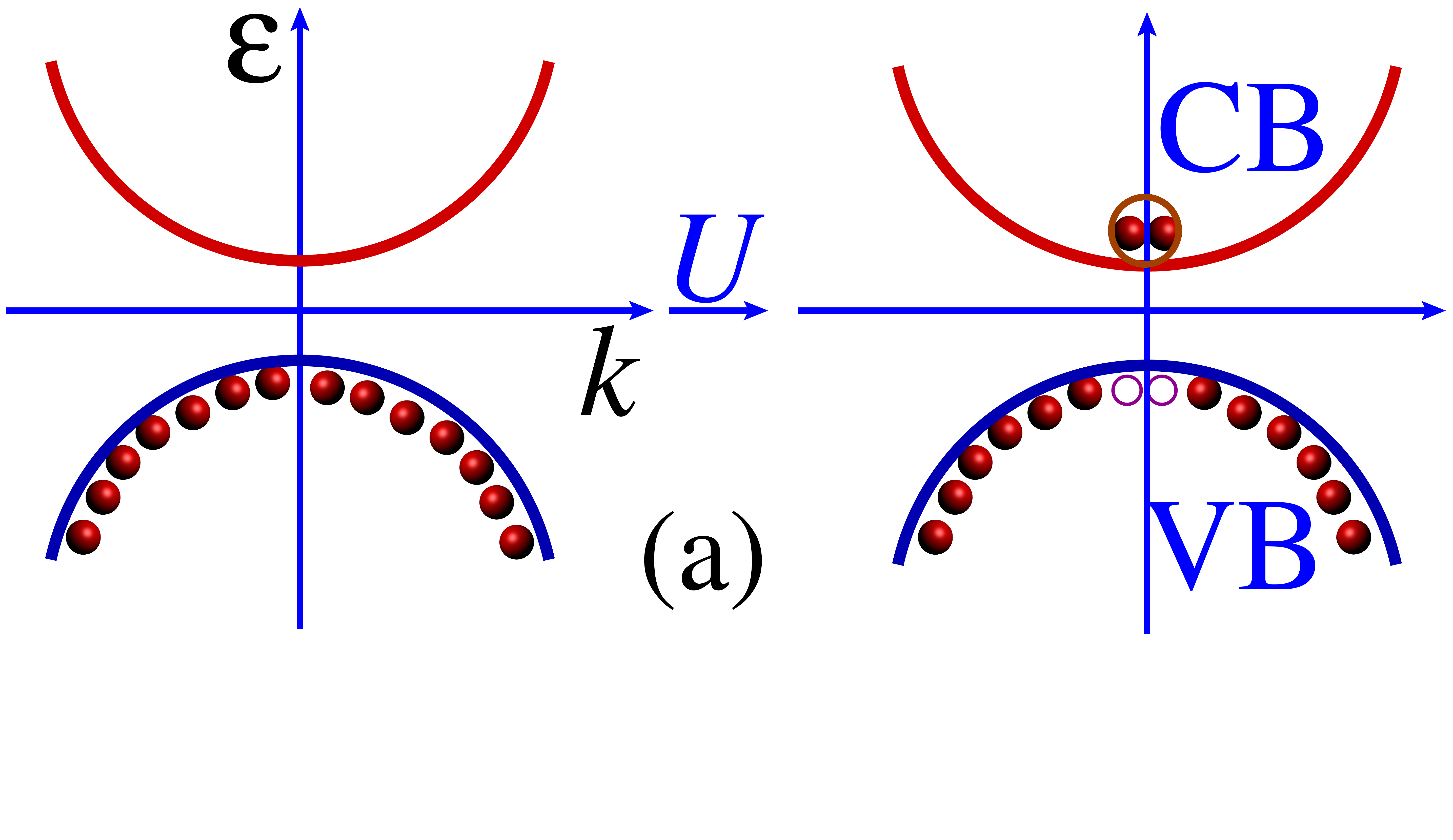}
     \includegraphics[width=\myhalffig]{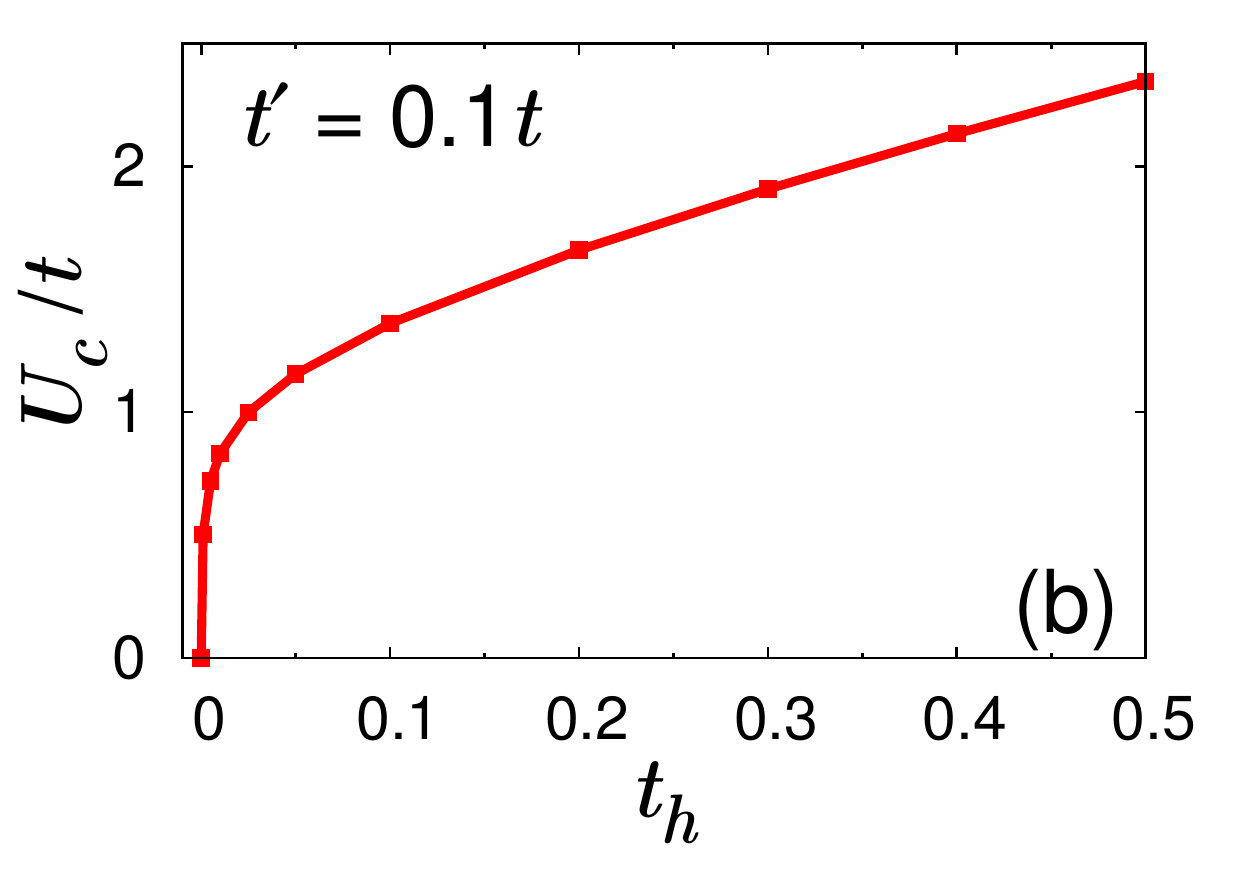} }
  \caption{(a) Schematic of Cooper instability. (b) The critical value  $U_c$ that induces Cooper instability. }
  \label{fig:Uc}
\end{figure}

Our proposal for the realization of a two dimensional spin-$\half$ fermionic superfluid state hinges on a
bilayer band insulator. The configuration consists of two layers $A$ and $B$, both of which have the same
lattice structure (such as a square or triangular lattice) and a 2D Brilliouin zone (see \Fig{fig:scheme}). The crucial ingredient is that the in-plane energy dispersion in the two layers are of
{\em opposite} sign, i.~e., $\varepsilon_A(\bk) = -\varepsilon_B(\bk) = \varepsilon(\bk)$ for all $\bk$ in
the Brilliouin zone. Inter-layer hopping is described by a hybridization function $h(\bk)$ which is such that
$h(\bk)$ and $\varepsilon(\bk)$ never vanish simultaneously. The kinetic energy of the system is thus described by
$ {\cal H}_{K}  = \sum_{\bk \sigma} \varepsilon(\bk) \left(a^\dagger_{\bk \sigma} a_{\bk \sigma} 
- b^\dagger_{\bk \sigma} b_{\bk \sigma} \right) + \left( h(\bk) a^\dagger_{\bk \sigma} b_{\bk \sigma}
+ \mbox{h.~c.} \right)$,
where $a$-s and $b$-s are spin-$\half$ fermion operators corresponding to $A$ and $B$ layers respectively
($\sigma=\uparrow,\downarrow$ is the spin). This leads to two (spin degenerate) bands
\beq\label{eqn:DiagKinEnergy}
{\cal H}_{K} = \sum_{\bk \sigma} e(\bk) \left(c^{\dagger}_{\bk \sigma} c_{\bk \sigma} - d^\dagger_{\bk \sigma} d_{\bk \sigma} \right) ,
\eeq
where $c$ and $d$ are respectively ``conduction'' and ``valance'' band fermion operators, with
$e(\bk) = \sqrt{\varepsilon(\bk)^2 + |h(\bk)|^2}$. With a fermion density of one particle per site on both
$A$ and $B$ layers, the ground state is the filled valance band, i.~e., a band insulator. We now introduce a local 
{\em attractive} interaction with strength $U$ (which may be tuned by a Feshbach resonance\mycite{Chin2010}) as
\beq\label{eqn:HubbardU}
\cH_U = -U \sum_{i} \left( a^\dagger_{i \uparrow}a^\dagger_{i \downarrow} a_{i \downarrow} a_{i \uparrow} +  b^\dagger_{i \uparrow}b^\dagger_{i \downarrow} b_{i \downarrow} b_{i \uparrow} \right).
\eeq
We show that a superfluid can be generated by starting from the band insulator ($U \approx 0$) and adiabatically
increasing the magnitude of the attractive interaction. To be specific, we choose a particular model to illustrate
the idea (see below for a possible laboratory realization of this model).  The bilayer system has both $A$ and $B$
layers which are square lattices (of unit lattice spacing) with nearest ($t$) and next nearest $(t')$ hopping
($t' = t/10$ throughout) such that $\varepsilon_A(\bk) = -\varepsilon_B(\bk) = \varepsilon(\bk)
= -2 t (\cos{k_x} + \cos{k_y})-4t' \cos{k_x} \cos{k_y}$ . The hybridization function $h(\bk) = -t_h$ captures the
hopping from adjacent $A$ and $B$ sites.  The resulting band structure is of the form in \eqn{eqn:DiagKinEnergy} and
has an energy gap $\upepsilon_g = 2t_h$.  The band insulator obtained with one particle per site per layer has a
Cooper instability\mycite{Cooper1956} at a non-zero critical value $U_c$ of $U$ unlike in a metal where $U_c$ is zero.
We find that $\frac{1}{U_c} \approx \frac{1}{N} \sum_{\bk} \frac{1}{e(\bk)} \sim \frac{1}{t} \ln{\frac{t}{\upepsilon_g}}$
(see \Fig{fig:Uc}), where $N$ is the number of sites per layer. The instability owes to the fact that at sufficiently large $U$,
it becomes feasible for a pair of fermions of opposite spin to be promoted to conduction band where they can ``sample''
the attractive interaction (see \Fig{fig:Uc}(a)), eventually forming a bound state.  We find that for $U \gtrsim U_c$,
the binding energy of the pair goes as $(U-U_c)$, which may be contrasted with exponentially small value usually
found\mycite{Cooper1956} in a system with a Fermi surface.  The physics of such strong binding owes to the
modification of the density of states at the band edges engendered by the hybridization. The resulting joint density of states
of particle-hole excitations is strongly enhanced (see \Fig{fig:scheme}), $g(\epsilon)\sim 1/\sqrt{\epsilon - \upepsilon_g}$,
and it is this large enhancement that provides for the strong binding as in other contexts.\mycite{Vyasanakere2011a}
Consequently we expect the system to also possess high transition temperatures making it attractive for experimental realization
of an optical lattice superfluid.

We now study the properties of the lattice superfluid state using functional integral techniques.\mycite{SaDeMelo1993,Dupuis2004}, by introducing the action
\beq
\mylabel{eqn:Action}
\begin{split}
  & \cS[\psi]  =  \sum_{k,\alpha} \psi^\star_{\alpha \sigma}(k)(-\mbox{G}_0^{-1}(k)) \psi_{\alpha \sigma}(k) \\
  + & \sum_k \left(h(\bk) \psi^\star_{1 \sigma}(k) \psi_{\bar{1} \sigma}(k) + \mbox{g.c.} \right)
- \frac{U}{\upbeta N} \sum_{q,\alpha} P^\star_\alpha(q) P_\alpha(q) ,
\end{split}
\eeq
where $-\mbox{G}_0^{-1}(k)= (-ik_n + \alpha \varepsilon(\bk) - \mu)$, $\mu$ is the chemical potential, $k = (ik_n, \bk)$, $(iq_\ell , \bq)$ 
with $ik_n(iq_\ell) = (2n +1)\pi/\upbeta \left(2 \ell \pi/\upbeta \right)$ being Fermi(Bose) Matsubara frequencies, and $\upbeta = 1/T$
is the inverse temperature. We have introduced Grassmann numbers $\psi_{\alpha \sigma}$ where the flavour label $\alpha=\pm 1$ stands, 
respectively, for $A$ and $B$ layers, and $P^\star_\alpha(q) = \sum_k \psi^\star_{\alpha \uparrow}(q+k) \psi^\star_{\alpha \downarrow}(-k)$.

\begin{figure}
  \centering 
  \centerline{ \includegraphics[width=\myhalffig]{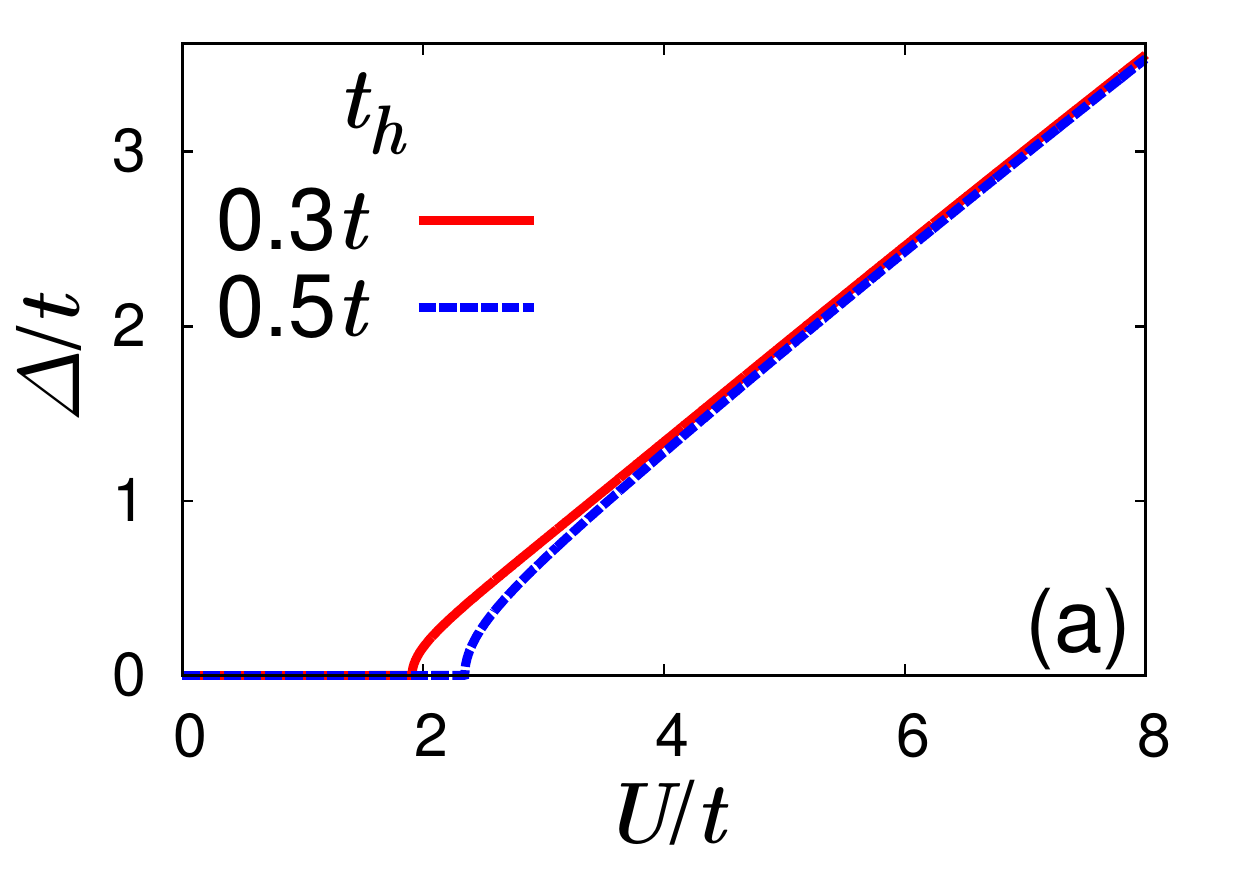}   \includegraphics[width=\myhalffig]{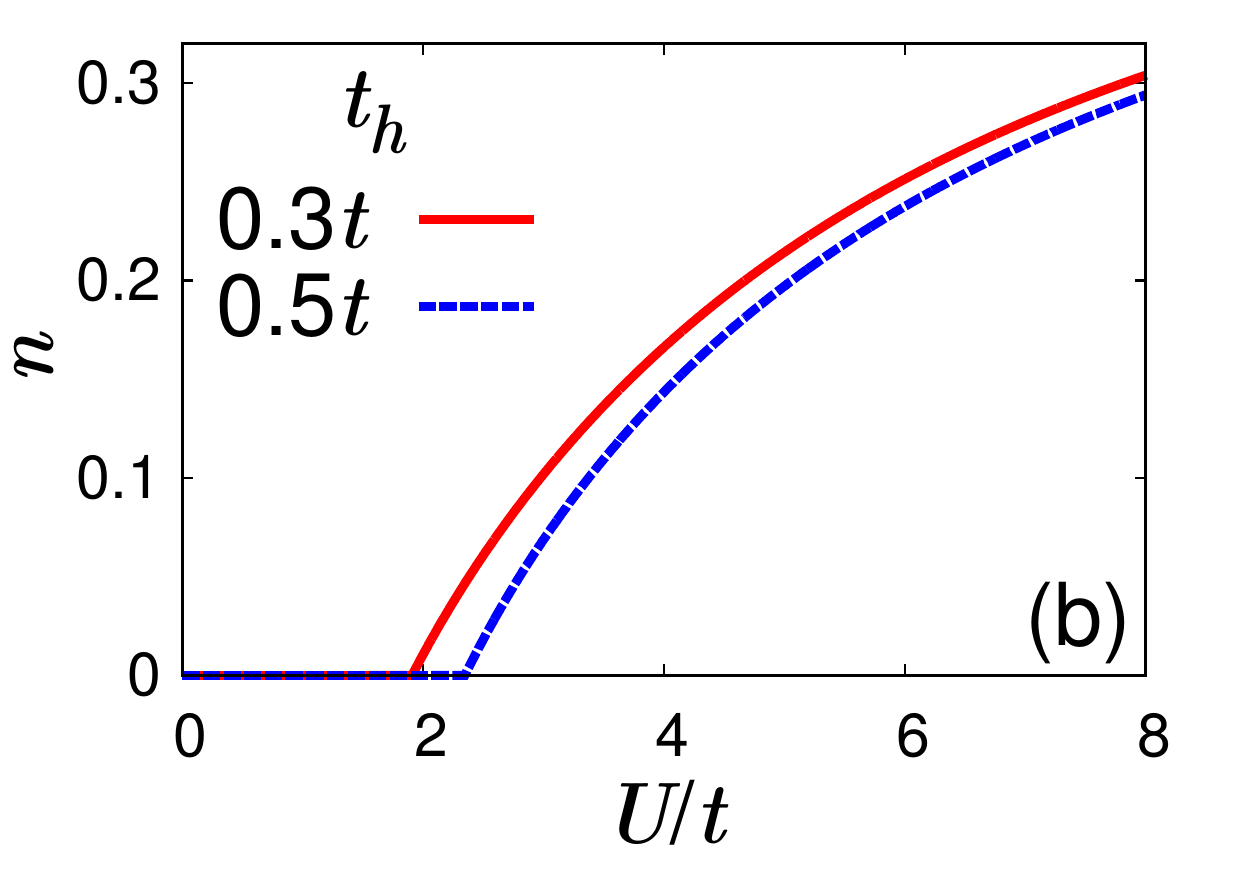} }
  \caption{Dependence of zero temperature {\bf (a)} superfluid order parameter $\varDelta$ and {\bf (b)} ``carrier density'' $n$ on $U$.}
  \label{fig:DeltaandN}
\end{figure}

Possibility of a superfluid state is investigated by introducing Hubbard-Stratanovich pair fields $\Delta_\alpha(q)$ to decouple 
the interaction term\cite{SM} in \eqn{eqn:Action}. The fermions are then integrated out to obtain an action $\cS[\Delta]$ solely for
$\Delta_\alpha(q)$. The uniform saddle point $\Delta^{C}_\alpha(q) = \varDelta \, \delta_{q,0}$, where $\varDelta$ is the
superfluid order parameter, gives the gap equation
\beq
\mylabel{eqn:Gap}
\frac{1}{U} = \frac{1}{N} \sum_{\bk} \frac{\tanh{\frac{\upbeta E(\bk)}{2}}}{2 E(\bk)} ,
\eeq
where $ E(\bk) = \sqrt{e(\bk)^2 + \varDelta^2}$. The model we consider here has particle hole symmetry which forces $\mu = 0$
when the occupancy is one fermion per site; thus a separate number equation to determine $\mu$ is obviated. \Fig{fig:DeltaandN}(a)
shows the evolution of zero temperature ground state with increasing $U$. At $U_c$, a quantum phase transition occurs ushering in a
superfluid state where $\varDelta$ behaves as $\sqrt{U-U_c}$ and monotonically increases with increasing $U$.
The superfluidity arises from promotion of fermions to the conduction band; indeed,
$n = \frac{1}{N}\sum_{\bk} \mean{c^\dagger_{\bk \sigma} c_{\bk \sigma}}$ 
(see \eqn{eqn:DiagKinEnergy}), the number of fermions promoted to the conduction band  increases from zero at $U_c$ as $n \sim (U - U_c)$.

Having established the superfluid ground state and its physical underpinnings, the natural question is regarding the magnitude of the 
transition temperature of the superfluid obtained. Two effects that destroy superfluidity are pair breaking and phase fluctuations. 
Indeed, long wavelength phase fluctuations render our 2D superfluid lacking in true long range order at finite temperatures, i.~e., 
in the Kosterliz-Thouless phase. The temperature scale of pair-breaking $T_{\varDelta}$ is set by the lowest temperature at which 
the saddle point value of $\varDelta$ vanishes, and can be obtained by solving for the temperature in the gap equation (\eqn{eqn:Gap}) 
with $\varDelta=0$. To investigate the role of phase fluctuations below $T_{\varDelta}$ and estimate the transition temperature, 
we study the fluctuations at the Gaussian level\cite{SM} by expanding the action $\cS[\Delta]$ about the saddle
point with $\Delta_\alpha(q) = \Delta_\alpha^C(q) + \varDelta( \zeta_\alpha(q) + i \theta_\alpha(q) )$ where $\zeta_\alpha(q)$ and $\theta_\alpha(q)$ are
real fields that represent, respectively, the amplitude and phase fluctuations in each layer. The fluctuations in each layer are
coupled; a more natural ``normal mode'' description is in terms of symmetric and anti-symmetric linear combinations of these
modes. For example, there are two phase modes -- the symmetric mode $\theta_s(q) \sim (\theta_+(q) + \theta_-(q))$, and the anti-symmetric
mode $\theta_a(q) \sim (\theta_+(q) - \theta_-(q))$, and there are two amplitude modes with similar definition.

\begin{figure}
  \centering 
  \centerline{ \includegraphics[width=\myhalffig]{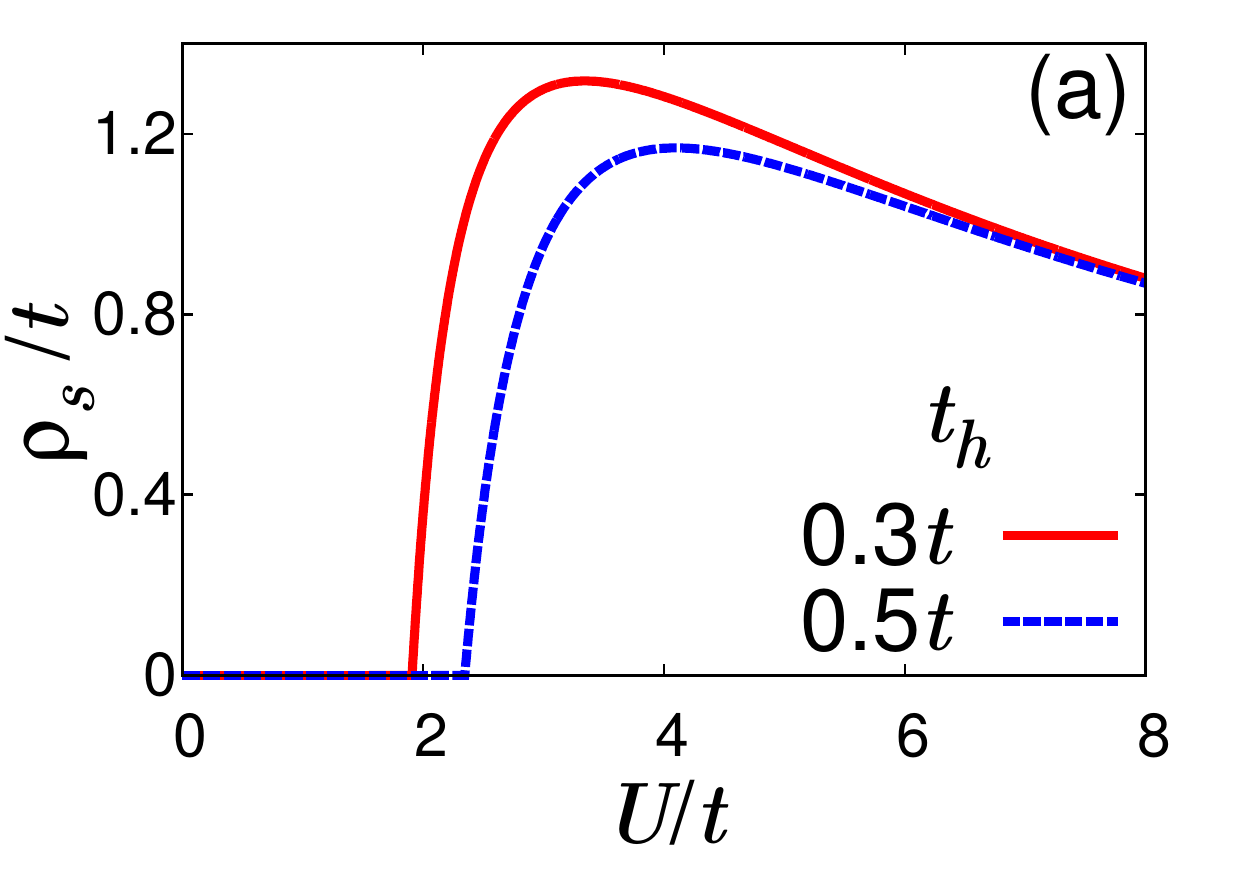} 
    \includegraphics[width=\myhalffig]{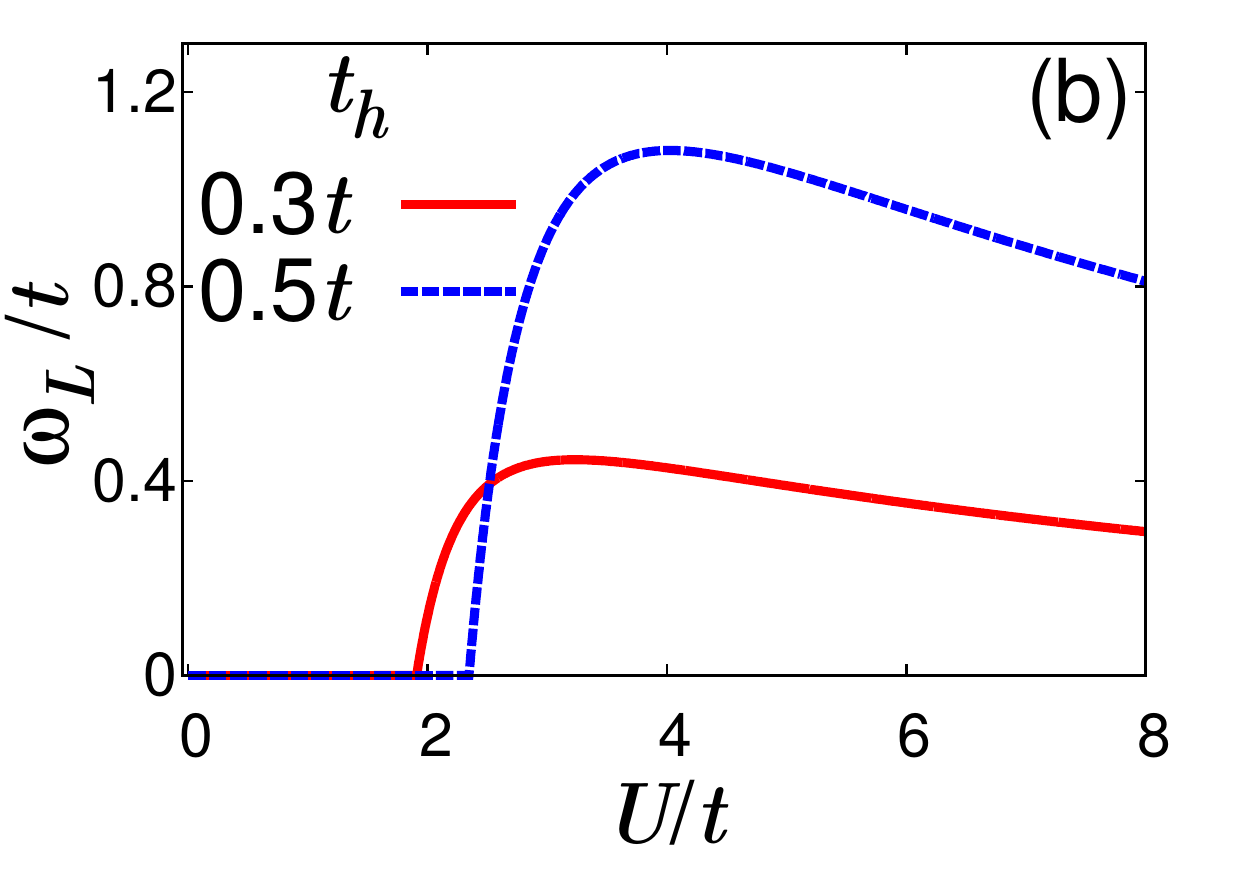} }
  \caption{Evolution of {\bf (a)} superfluid density $\rho_s$ and {\bf (b)} 
    Leggett mode gap $\omega_L$ as a function of $U$ at zero temperature.}
  \label{fig:SFD}
\end{figure}

We find that both the amplitude modes are gapped, while the symmetric phase mode is gapless and the anti-symmetric phase mode is gapped. 
Interestingly, the gapped anti-symmetric phase mode is analogous to the Leggett mode in multi-band superconductors.\mycite{Leggett1966,Zhao2006}
We obtain the following effective action for the phase modes by integrating out the amplitude modes 
$ \cS[\theta_s,\theta_a] =$ $\int_0^\upbeta \D{\tau} \int \D{^2\br} \left[\left(\kappa_s \left( \frac{\dou \theta_s}{\dou \tau} \right)^2 \right.
\right.$ $ \left.  + \rho_s \left(\frac{\dou \theta_s}{\dou \br} \right)^2 \right) + $ $ \left( \kappa_a \left( \frac{\dou \theta_a}{\dou \tau}
\right)^2  + \right. $ $\left. \left. \rho_a \left(\frac{\dou \theta_a}{\dou \br} \right)^2  + \omega_L \theta_a^2\right) \right]$,
where $\tau$ is the imaginary time, $\br$ is the position on the 2D plane, $\kappa$-s, $\rho$-s and $\omega_L$ are determined by the
saddle point solution. The most important parameters\cite{SM} in this action are the phase stiffness of the symmetric phase mode $\rho_s$,
which is the superfluid density, and $\omega_L$, the gap (or mass) associated with the anti-symmetric Leggett mode. 
\Fig{fig:SFD}(a) shows the dependence of the zero-temperature superfluid density $\rho_s$ on $U$. For $U \gtrsim U_c$, $\rho_s \sim U-U_c$,
and has the same behaviour as the number of fermions excited to the conduction band. With increase of $U$, $\rho_s$ attains a maximum,
and suffers a fall at larger values of $U$. For $U \gg t$, we find that $\rho_s \sim \frac{t^2}{U}$. 
This owes to the fact that the system undergoes a ``BCS-BEC'' crossover with increasing $U$, and $\rho_s$ is determined by the
hopping amplitude of the bosonic fermion pair at large $U$ which is $\sim t^2/U$. The variation of $\omega_L$ with $U$ is shown
in \Fig{fig:SFD}(b), the key point to be noted is that in the regime where $\rho_s$ is largest, the Leggett mode has 
a large gap, and does not participate in the low energy physics.

\begin{figure}
  \centering 
  \centerline{  \includegraphics[width=\myhalffig]{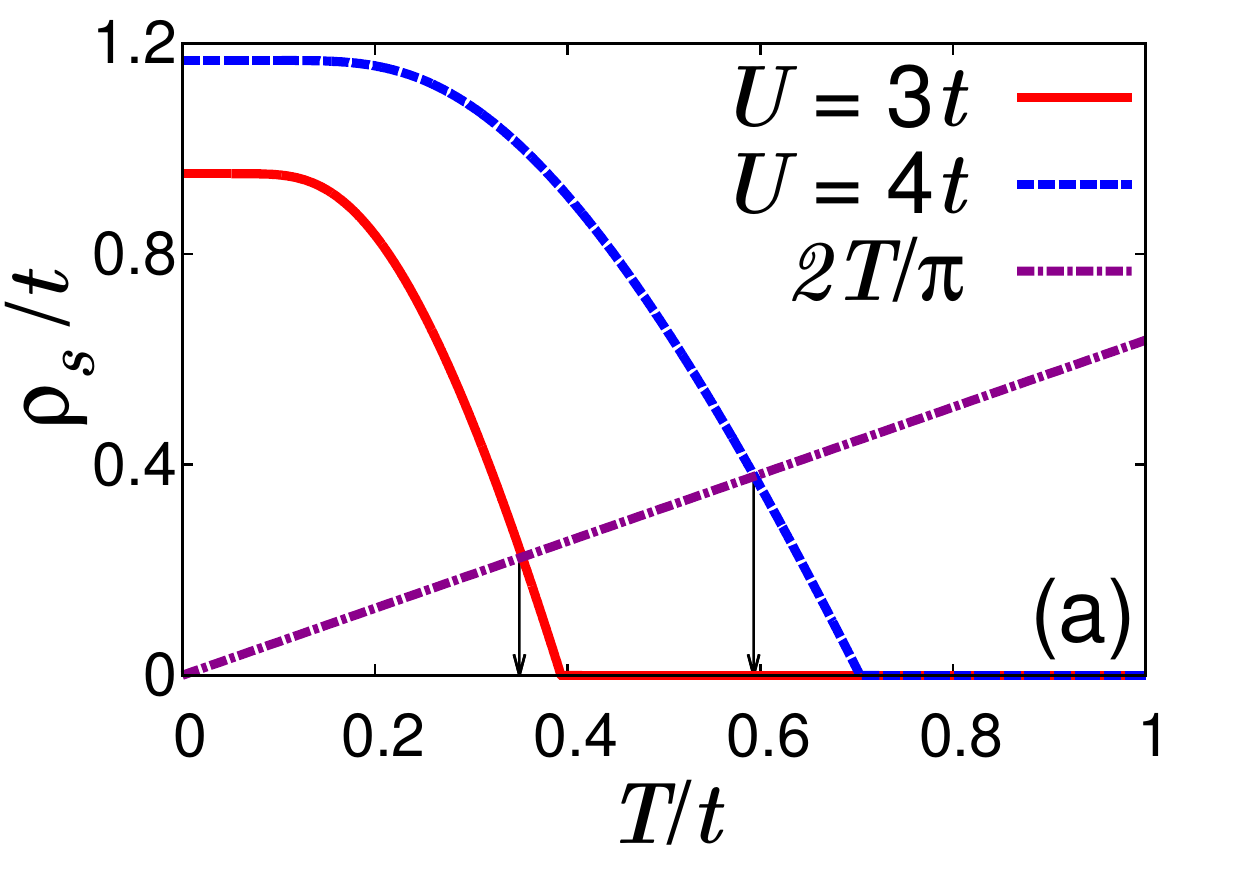}  \includegraphics[width=\myhalffig]{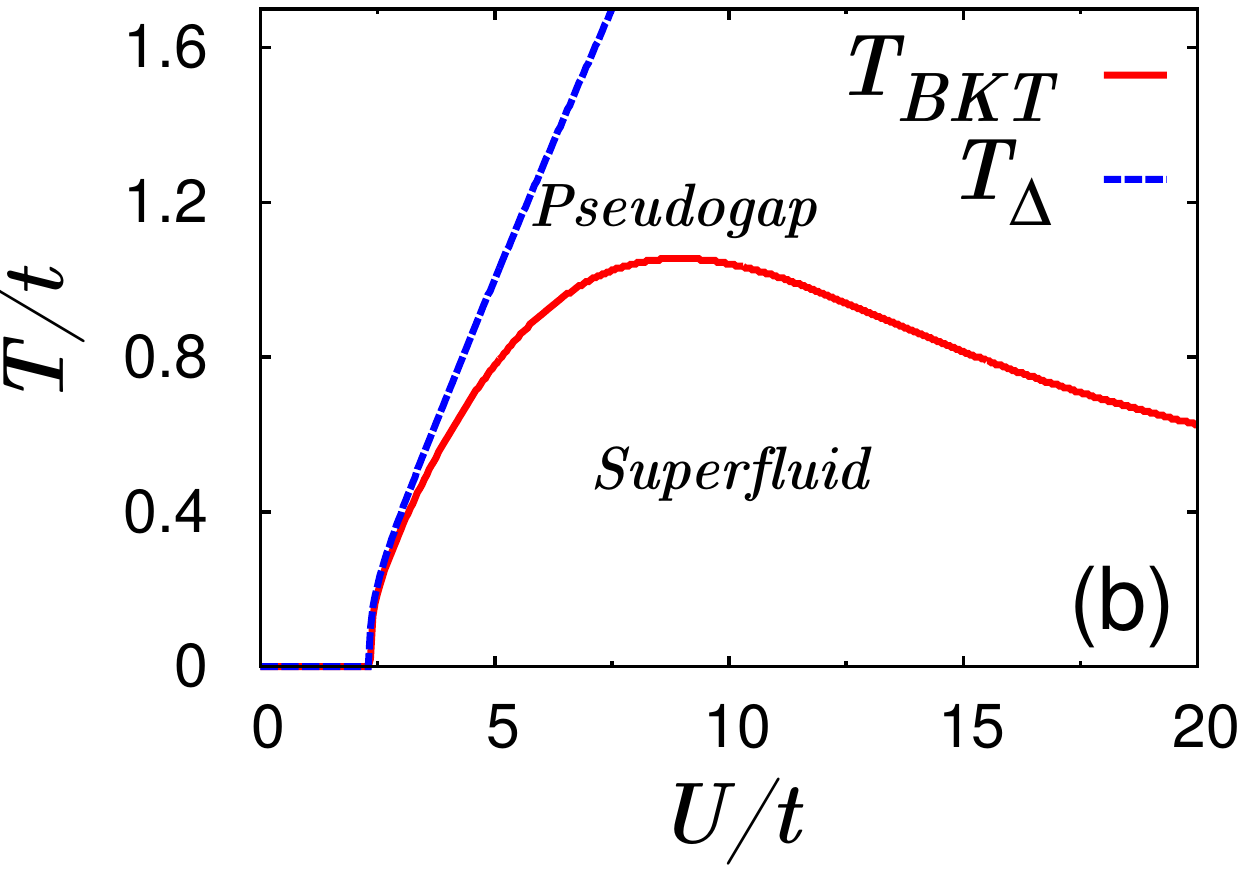} }
  \caption{{\bf (a)} Determination of $T_{BKT}$ from the temperature dependence of $\rho_s$.
  {\bf (b)} Dependence of $T_{\varDelta}$ and $T_{BKT}$ on $U$. Note the high transition temperature, and the large pseudogap regime.
  Here $t_h=0.5t$.}
  \label{fig:TBKT}
\end{figure}

The discussion above allows the estimation of the Kosterlitz-Thouless transition temperature $T_{BKT}$.
We obtain $\rho_s$ as a function of $T$ via our functional formulation using the saddle point value of
$\varDelta$. Using the relationship\mycite{Chaikin1995} that $\rho_s(T_{BKT}) = \frac{2 T_{BKT}}{\pi}$,
we arrive at the transition temperature (as shown in \Fig{fig:TBKT}(a)) plotted in \Fig{fig:TBKT}(b), 
which also shows the temperature $T_{\varDelta}$ associated with pair breaking obtained from \eqn{eqn:Gap}.
We see that the maximum value of $T_{BKT}$ is of the order of the lattice hopping amplitude $t$, and in
this sense we obtain {\em high temperature} superfluidity in the regime where the crossover to the BEC state
takes place.  The BCS side ($U \gtrsim U_c$) is also a robust superfluid owing to the enhancement obtained
by the divergent density of states.  The transition temperature of the system can also be affected by the
vortex core energies\mycite{Benfatto2007}, but  these effects will be unimportant in this system. Another
attractive aspect of this system is that one expects to see large pseudogap features even at high
temperatures (see \Fig{fig:TBKT}(b)), and thus interesting physics can be investigated in optical lattices 
even if the average entropy of the system is not small.

\begin{figure}
  \centering 
  \centerline{ \includegraphics[width=\myhalffig]{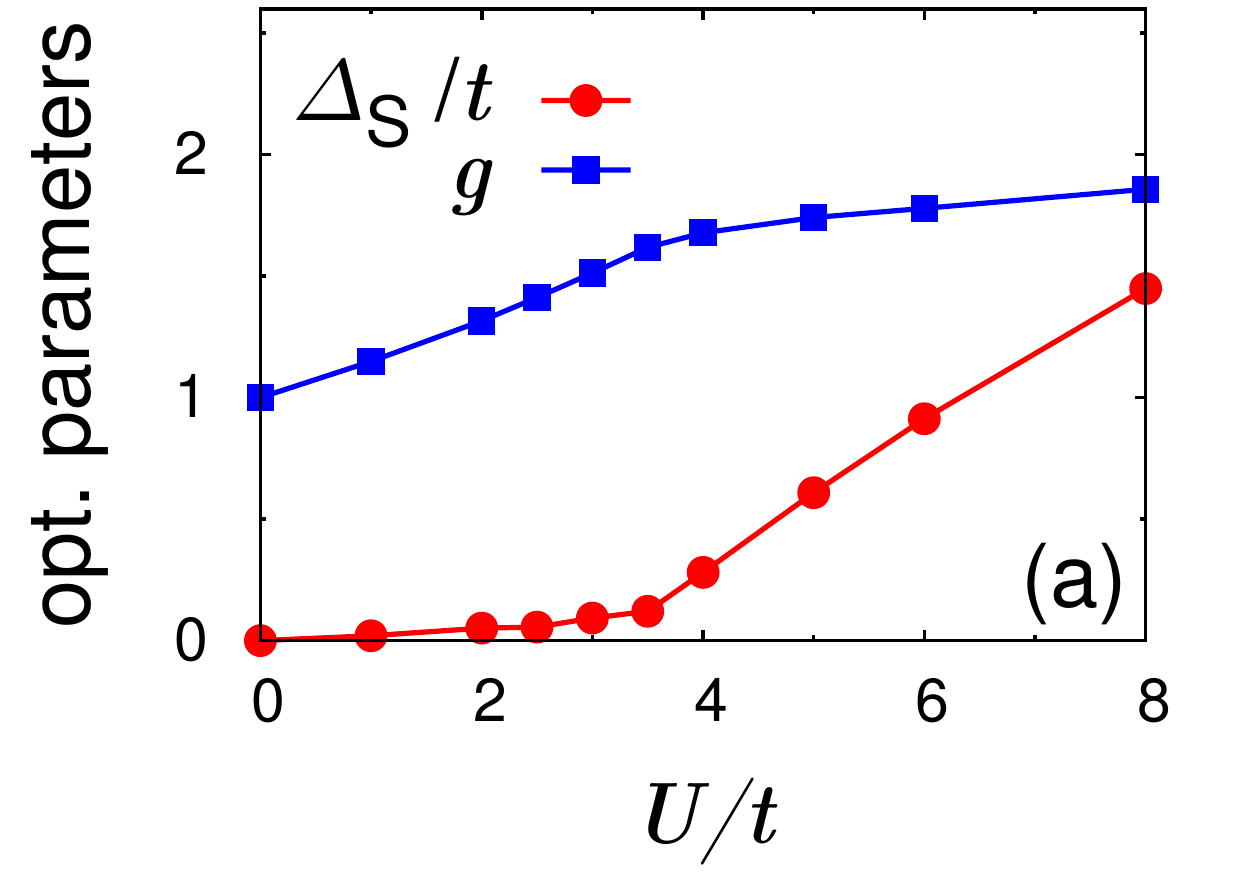}   \includegraphics[width=\myhalffig]{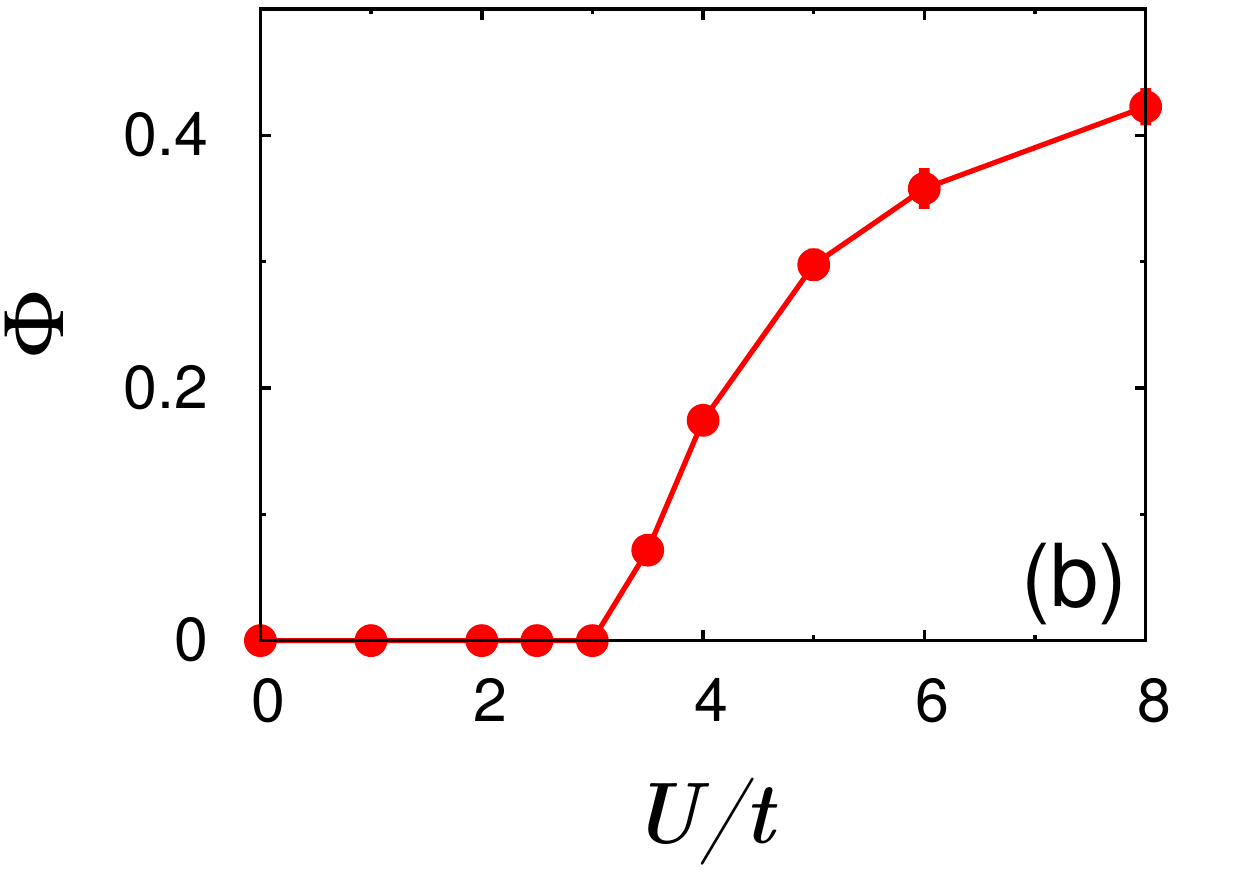} }
  \caption{Results of variational Monte Carlo calculations. Dependence of {\bf (a)}  variational parameters $g$ and $\varDelta_S$ and {\bf (b)} superfluid order parameter $\Phi$, on $U$. $t_h=0.5t$.}
  \label{fig:VMC}
\end{figure}

The effect of quantum fluctuations are likely important due to the reduced dimensionality.\mycite{Paramekanti2000}
To ensure that quantum fluctuations only have a quantitative role, and to ensure that there are no competing orders
such as a CDW intervening, we conducted a detailed variational Monte-Carlo calculation of the ground state.\cite{SM} Our
variational ground state $\ket{\Psi} = g^{D}\ket{\varDelta_S,\varDelta_{CDW}}_{BCS}$ is constructed by introducing
both the superfluid pair order $\varDelta_S$, and a commensurate $(\pi,\pi)$ charge density wave order parameter
$\varDelta_{CDW}$, and obtaining the BCS state $\ket{\varDelta_S,\varDelta_{CDW}}_{BCS}$.
The Gutzwiller parameter $g$ ($>1$) that promotes double occupancy
($D$ is the operator that counts the number of doubly occupied sites)
introduces quantum fluctuations of the local phase. Two key results of our detailed study are: (i) For all values of
$U$ within the range considered here, the optimal value of $\varDelta_{CDW}$ is zero, i.~e., there is no competing order
that intervenes and hence the superfluid state is stable, (ii) quantum fluctuations do not change the qualitative
aspects of the results. Indeed, for the parameter values shown in \Fig{fig:VMC}(a), we find $U_c\approx 3.2$
is expectedly larger than the value of $2.3$ from the saddle point analysis. The variational parameter $\varDelta_S$
(\Fig{fig:VMC}(a)) and the superfluid order parameter $\Phi$ (\Fig{fig:VMC}(b)), which measures the amplitude
of injecting a pair at a large distance away from the point of its removal, has precisely the behaviour as expected
from the saddle point analysis.

\begin{figure}
  \centering 
  \centerline{ \includegraphics[width=5.0truecm]{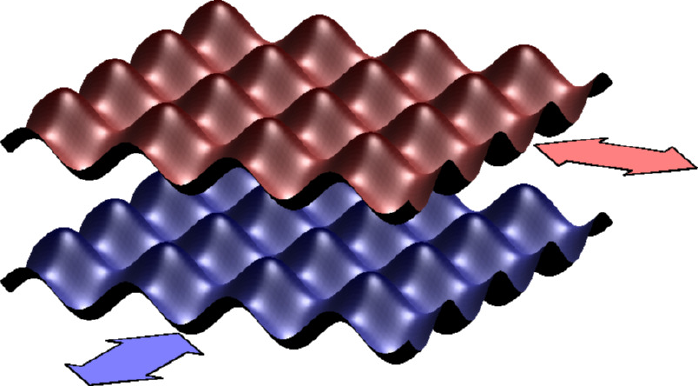}}
  \caption{Orthogonally shaken bilayer optical lattice. The top layer $A$ is shaken in the $x$ direction, while the bottom layer $B$ is shaken in the $y$ direction. By an appropriate choice of laser intensities, amplitude and frequency of the shakes, the band dispersion in the layers can be made to have opposite signs to each other. The layer hybridization can be controlled by the distance between the layers.}
  \label{fig:Proposal}
\end{figure}

Our proposed scheme can be realized by an ``orthogonally shaken bilayer'' depicted in \Fig{fig:Proposal}. It was argued in
\myonlinecite{Eckardt2005} that on introducing a shake of the optical lattice, the amplitude and {\em sign} of the hopping
can be controlled. It was shown that if $K$ (an energy scale) and $\nu$ are respectively the amplitude and frequency of the
shake, the effective hopping amplitude $t_{eff} = t J_0(\frac{K}{\nu})$ where $J_0$ is the Bessel function and $t$ is the
hopping in the absence of the shake. This phenomenon has not only been observed experimentally\mycite{Lignier2007}, but has
been recently used to study many interesting quantum phases\mycite{Eckardt2010,Struck2011} with further proposals for the
generation of topological insulators.\mycite{Hauke2012,Struck2012}

Our proposed experimental set up consists of two adjacent optical square lattices. The top layer $A$ is obtained by
interfering two sets of counter propagating laser beams in the $x$ and $y$ directions; the $x$ and $y$ beams are
non-interfering. The relative phase of the two $x$ laser beams is modulated so as to obtain a shake, and intensity
of the $x$ laser beams and amplitude of the modulation can be chosen such that $-t^A_x = t^A_y=t$, i.~e., the hopping
along the $x$ direction has an opposite sign to that in the $y$ direction. In the layer $B$, the beams along
the $y$-direction are shaken so that $t^A_x = -t^B_y = t$. This provides a realization of a system with $\varepsilon_A(\bk) =
-\varepsilon_B(\bk)$. The hybridization of the two layers can be controlled by the distance between the two layers.
This can be achieved by using vertically confining beams as in ref.~\myonlinecite{Gemelke2009}, and creating the two layers by
``optical copying''. Optical copying will entail splitting the $x$ and $y$ laser beams of the $A$ layer and focusing the split
beams just below the $A$ layer to produce the $B$ layer such that the $x$ beam of the $A$ layer plays the role of the $y$ beam
of the $B$ layer, and so on. This laboratory realization of our proposal may require optics techniques that have been used in
the making of quantum gas microscope.\mycite{Bakr2010} The confining trap potential is to be designed such that a large region
near the trap center will be in a band insulating state, with the excess entropy trapped in regions at the periphery.
Tuning of the attractive interactions should now drive the central band insulating region to the superfluid state.
We hope that this work stimulates experimental research on realizing such a bilayer band insulator system,
even by routes other than our proposal.

YP and AM thank CSIR and IISc-CPDF respectively for support. VBS is
grateful to DST(Ramanujan Grant) and DAE-SRC for generous support.
The authors thank Tilman Esslinger for discussions and suggestions
regarding experimental realization of the proposed model, and Jayantha
Vyasanakere for discussions/comments, and Arun Paramekanti for
comments on the manuscript.

\bibliography{bibliography_inssc}

\end{document}